\documentclass[aps,prd,onecolumn,preprintnumbers,groupedaddress,showpacs,nofootinbib,amssymb]{revtex4}
\usepackage{graphicx,color}
\usepackage{amsmath}
\usepackage{amssymb}
\usepackage{amsfonts}
\usepackage{bm}
\usepackage{cancel}
\usepackage{graphicx}
\usepackage{epsf}
\usepackage{bm}
\usepackage{amsmath}
\usepackage{amsfonts}
\usepackage{amssymb}
\usepackage{color}
\usepackage{epstopdf}

\allowdisplaybreaks[4]

\newcommand{\bea}{\begin{eqnarray}}
\newcommand{\eea}{\end{eqnarray}}

\newcommand{\ba}{\begin{eqnarray}}
\newcommand{\ea}{\end{eqnarray}}
\newcommand{\be}{\begin{equation}}
\newcommand{\ee}{\end{equation}}

\begin{document}

\title{Cosmological constant and renormalization of gravity}
\author{Shin'ichi Nojiri$^{1,2,3}$}

\affiliation{
$^1$ Department of Physics,
Nagoya University, Nagoya 464-8602, Japan \\
$^2$ Kobayashi-Maskawa Institute for the Origin of Particles and the Universe, 
Nagoya University, Nagoya 464-8602, Japan \\
$^3$ KEK Theory Center, High Energy Accelerator Research Organization (KEK),
Oho 1-1, Tsukuba, Ibaraki 305-0801, Japan}

\date{\today}
\begin{abstract}
In arXiv:1601.02203 and arXiv:1702.07063, we have proposed 
a topological model with a simple Lagrangian density and have tried 
to solve one of the cosmological constant problems. 
The Lagrangian density is the BRS exact and therefore the model can be 
regarded as a topological theory. 
In this model, the divergence of the vacuum energy coming from the quantum 
corrections from matters can be absorbed into the redefinition of the scalar field. 
In this paper, we consider the extension of the model in order to apply the mechanism to 
other kinds of divergences coming from the quantum correction and consider 
the cosmology in an extended model. \end{abstract}

\maketitle

\section{Introduction}

By the recent cosmological observations, we now believe the accelerating expansion of 
the present universe, whose simplest model may be given by 
a cosmological term with a small cosmological constant. 
We also know that the quantum correction coming from the contributions 
from matters to the vacuum energy, which may be identified with 
the cosmological constant, terribly diverges and we need the very finely tuned 
counterterm to cancel the divergence. 
For the discussion about the small but non-vanishing vaccum energy, 
see \cite{Burgess:2013ara} for example. 
In order to solve the problem of the large quantum corrections to the vacuum 
energy, the unimodular gravity theories \cite{Anderson:1971pn,
Buchmuller:1988wx,Buchmuller:1988yn, Henneaux:1989zc,Unruh:1988in,
Ng:1990xz,Finkelstein:2000pg,Alvarez:2005iy,Alvarez:2006uu,
Abbassi:2007bq,Ellis:2010uc,Jain:2012cw,
Singh:2012sx,Kluson:2014esa,Padilla:2014yea,Barcelo:2014mua,
Barcelo:2014qva,Burger:2015kie,
Alvarez:2015sba,Jain:2012gc,Jain:2011jc,Cho:2014taa,Basak:2015swx,
Gao:2014nia,Eichhorn:2015bna,
Saltas:2014cta,Nojiri:2015sfd} have been proposed and discussed. 
Other scenarios like the sequestering mechanism have been also proposed 
\cite{Kaloper:2013zca,Kaloper:2014dqa,Kaloper:2015jra, 
Batra:2008cc,Shaw:2010pq,Barrow:2010xt,Carballo-Rubio:2015kaa}. 
In \cite{Nojiri:2016mlb}, we have proposed a new model which could be 
regarded with a topological field theory and desicussed the cosmology in 
the model \cite{Mori:2017dhe}. 
We should note that by the quantum corrections from the matter, 
the following terms, besides the cosmological constant, are generated, 
\begin{equation}
\label{A3}
\mathcal{L}_\mathrm{qc} = \alpha R + \beta R^2 
+ \gamma R_{\mu\nu} R^{\mu\nu} 
+ \delta R_{\mu\nu\rho\sigma} R^{\mu\nu\rho\sigma} \, .
\end{equation}
Here the coefficient $\alpha$ diverges quadratically and $\beta$, $\gamma$, 
and $\delta$ diverge logarithmically without the cut-off scale. 
If we further include the quantum corrections from the graviton, 
infinite numbers of diverging quantum corrections appear. 
In order to solve these problems of the divergences and the 
renormalizations, we extend the model in \cite{Nojiri:2016mlb} and discuss 
the cosmology given by the extended models. 
In the next section, we review on the topological model 
for the cosological constant based on \cite{Nojiri:2016mlb}. 
In Section \ref{Sec3}, we extend the toplogical model and consider more 
general divergences. In Section \ref{Sec4}, we consider the cosmology 
given by the extended topological model. 
The last section is devoted for summary and discussions. 

\section{Topological Model for Cosmological Constant Problem \label{Sec2}}

The action of the model in \cite{Nojiri:2016mlb} has the following form, 
\begin{equation}
\label{CCC7} 
S' = \int d^4 x \sqrt{-g} \left\{ \mathcal{L}_\mathrm{gravity} 
 - \lambda + \partial_\mu \lambda \partial^\mu \varphi  
 - \partial_\mu b \partial^\mu c \right\} 
+ S_\mathrm{matter} \, .
\end{equation}
Although $\lambda$ and $\varphi$ are ordinary scalar fields but  
$b$ and $c$ are fermionic (Grassmann odd) scalar fields and 
we regard that $b$ is an anti-ghost field and $c$ with a ghost field.\footnote{
The action without $c$ and $b$ has been proposed  
in \cite{Shlaer:2014gna} in order to solve the problem of time. }\footnote{
The cosmological perturbation in the model motivated in the model 
(\ref{CCC7}) has been investigated in \cite{Saitou:2017zyo}. } 
In (\ref{CCC7}), $S_\mathrm{matter}$ is the action of matter and 
$\mathcal{L}_\mathrm{gravity}$ is the Lagrangian density of 
arbitrary gravity.  
There does not appear any parameter in the action (\ref{CCC7}) 
except in the parts of $S_\mathrm{matter}$ and 
$\mathcal{L}_\mathrm{gravity}$. 

By separating the gravity Lagrangian density $\mathcal{L}_\mathrm{gravity}$ 
into the sum of some constant $\Lambda$, which may include the large 
quantum corrections from matter, and other part 
$\mathcal{L}_\mathrm{gravity}^{(0)}$,  
$\mathcal{L}_\mathrm{gravity} = \mathcal{L}_\mathrm{gravity}^{(0)} 
 - \Lambda$, we redefine the scalar field $\lambda$ by 
 $\lambda \to \lambda - \Lambda$. 
The obtained action has the following form, 
\begin{equation}
\label{CCC7R} 
S' = \int d^4 x \sqrt{-g} \left\{ \mathcal{L}_\mathrm{gravity}^{(0)}
 - \lambda + \partial_\mu \lambda \partial^\mu \varphi  
 - \partial_\mu b \partial^\mu c \right\} 
+ S_\mathrm{matter} \, .
\end{equation}
Note that the action (\ref{CCC7R}) does not include the cosmological 
constant $\Lambda$ and therefore the constant $\Lambda$ never 
affects the dynamics in the model. 
This tells that the large quantum corrections from the matters 
can be tuned to vanish. 

As shown in \cite{Nojiri:2016mlb}, there appear the ghosts, which generate 
the negative norm states in the quantum theory, in the model (\ref{CCC7}). 
The existence of the negative norm states makes the so-called Copenhagen 
interpretation invalid and therefore the model becomes inconsistent. 
The negative norm states, however, can be eliminated by defining 
the physical states which are annihilated by the BRS charge 
\cite{Becchi:1975nq}. 
We can find that the action (\ref{CCC7}) is invariant under the infinite 
numbers of the BRS transformation, 
\begin{equation}
\label{CCC8BR}
\delta \lambda = \delta c = 0\, , \quad 
\delta \varphi = \epsilon c \, , \quad 
\delta b = \epsilon \left( \lambda - \lambda_0 \right)\, .
\end{equation}
Here $\epsilon$ is a fermionic (Grassmann odd) parameter and 
$\lambda_0$ should satisfy the following equation by putting $\lambda=\lambda_0$, 
\begin{equation}
\label{lambda0}
0 = \nabla^\mu \partial_\mu \lambda\, ,
\end{equation}
which is obtained by the variation of the action (\ref{CCC7}) 
with respect to $\varphi$.\footnote{
The existence of the BRS transformation where $\lambda_0$ satisfies
Eq.~(\ref{lambda0}) was pointed out by R. Saitou.} 
If the physical states are defined as the states invariant under the BRS 
transformation in (\ref{CCC8BR}), the negative norm states 
can be eliminated by the Kugo-Ojima mechanism in the gauge theory 
\cite{Kugo:1977zq,Kugo:1979gm}.\footnote{
We can assign the ghost number, which is conserved, $1$ for $c$ and $-1$ 
for $b$ and $\epsilon$. 
The four scalar fields $\lambda$, $\varphi$, $b$, and $c$ are called 
a quartet \cite{Kugo:1977zq,Kugo:1979gm}}

Because Eq.~(\ref{CCC8BR}) tells that $\lambda - \lambda_0$ is 
given by the BRS transformation of the anti-ghost $b$ and therefore 
the vacuum expectation value of $\lambda - \lambda_0$ must vanish 
in the physical states. 
Therefore there occurs the spontaneous breakdown of the corresponding BRS 
symmetry in case that the vacuum expectation value of 
$\lambda - \lambda_0$ does not vanish. 
For the broken BRS symmetry, it is impossible to impose 
the physical state condition.  
It should be noted, however, there is one and only one unbroken 
BRS symmetry in the infinite numbers of the BRS symmetries 
in (\ref{CCC8BR}). 
The point is that Eq.~(\ref{lambda0}) is nothing but the field equation 
for $\lambda$. 
Because the real world is realized by one and only one solution of 
(\ref{lambda0}) for $\lambda$, one and only one $\lambda_0$ is chosen 
so that $\lambda=\lambda_0$ and therefore the  
corresponding BRS symmetry is not broken, which eliminates the 
negative norm states, which are the ghost states, and the unitarity is 
guaranteed. 
Although the quantum fluctuations are prohibited by the BRS symmetry, 
$\lambda_0$ can include the classical fluctuation as long as $\lambda_0$ 
satisfies the classical equation (\ref{lambda0}). 

We can regard the Lagrangian density in the action (\ref{CCC7}), 
\begin{equation}
\label{SCCP1}
\mathcal{L} =  - \lambda + \partial_\mu \lambda 
\partial^\mu \varphi  - \partial_\mu b \partial^\mu c \, ,
\end{equation}
as the Lagrangian density of a topological field theory 
\cite{Witten:1988ze}, where the Lagrangian density is given by 
the BRS transformation of some quantity. 
If we consider the model which only include the scalar field $\varphi$ 
but whose Lagrangian density identically vanishes, which tells that  
the action is trivially invariant under any transformation of $\varphi$.  
We may fix the gauge symmetry by imposing the following gauge 
condition, 
\begin{equation}
\label{CCC9}
1 + \nabla_\mu \partial^\mu \varphi = 0\, .
\end{equation}
By following the paper \cite{Kugo:1981hm}, we find the gauge-fixed 
Lagrangian is given by the BRS transformation (\ref{CCC8BR}) of 
$- b \left( 1 + \nabla_\mu \partial^\mu \varphi \right)$ and we obtain 
\begin{equation}
\label{SCCP2R}
\delta \left(- b \left( 1 + \nabla_\mu 
\partial^\mu \varphi \right) \right)
= \epsilon \left( - \left(\lambda - \lambda_0 \right)  
\left( 1 + \nabla_\mu \partial^\mu \varphi \right) 
+ b \nabla_\mu \partial^\mu c \right) 
= \epsilon \left( \mathcal{L} + \lambda_0 
+ \left(\mbox{total derivative terms}\right) 
\right)\, .
\end{equation}
Then we find the Lagrangian density (\ref{SCCP1}) is given by the BRS 
transformation of the quantity 
$- b \left( 1 + \nabla_\mu \partial^\mu \varphi \right)$ 
up to the total derivative terms if $\lambda_0=0$. 
The action is not given by the BRS transformation (\ref{CCC8BR}) with the non-vanishing 
$\lambda_0$, which could be a reason 
why the Lagrangian density  (\ref{SCCP1}) gives non-trivial and physically 
relevant contributions. 

\section{Extension of Topological Model \label{Sec3}}

The mechanism in the last sectioncan work for the divergences in (\ref{A3}) or more 
general divergences \cite{Mori:2017dhe}. 
In case that we include the divergences in (\ref{A3}), we may generalize 
the model in (\ref{SCCP1}) as follows, 
\begin{eqnarray}
\label{A4}
\mathcal{L} =&   - \Lambda - \lambda_{(\Lambda)} 
+ \left( \alpha  + \lambda_{(\alpha)} \right)R 
+ \left( \beta + \lambda_{(\beta)} \right) R^2 
+ \left( \gamma + \lambda_{(\gamma)} \right) R_{\mu\nu} R^{\mu\nu} 
+ \left( \delta + \lambda_{(\delta)} \right) 
R_{\mu\nu\rho\sigma} R^{\mu\nu\rho\sigma} \nonumber \\
& + \partial_\mu \lambda_{(\Lambda)} \partial^\mu \varphi_{(\Lambda)}  
 - \partial_\mu b_{(\Lambda)} \partial^\mu c_{(\Lambda)} 
+ \partial_\mu \lambda_{(\alpha)} \partial^\mu \varphi_{(\alpha)}  
 - \partial_\mu b_{(\alpha)} \partial^\mu c_{(\alpha)} \nonumber \\
& + \partial_\mu \lambda_{(\beta)} \partial^\mu \varphi_{(\beta)}  
 - \partial_\mu b_{(\beta)} \partial^\mu c_{(\beta)} 
+\partial_\mu \lambda_{(\gamma)} \partial^\mu \varphi_{(\gamma)}  
 - \partial_\mu b_{(\gamma)} \partial^\mu c_{(\gamma)} 
+ \partial_\mu \lambda_{(\delta)} \partial^\mu \varphi_{(\delta)}  
 - \partial_\mu b_{(\delta)} \partial^\mu c_{(\delta)} \, .
\end{eqnarray}
As in the case of the vacuum energy, the divergences are included in 
the coefficients $\Lambda$, $\alpha$, $\beta$, $\gamma$, and $\delta$ but 
if we shift the parameters 
$\lambda_{(\Lambda)}$, $\lambda_{(\alpha)}$, $\lambda_{(\beta)}$, 
$\lambda_{(\gamma)}$, and $\lambda_{(\delta)}$ as follows, 
\begin{equation}
\label{A5}
\lambda_{(\Lambda)} \to \lambda_{(\lambda)} - \Lambda\, , \quad 
\lambda_{(\alpha)} \to \lambda_{(\alpha)}  - \alpha\, , \quad 
\lambda_{(\beta)} \to \lambda_{(\beta)} - \beta\, , \quad 
\lambda_{(\gamma)} \to \lambda_{(\gamma)} - \gamma\, , \quad 
\lambda_{(\delta)} \to \lambda_{(\delta)} - \delta \, ,
\end{equation}
we can rewrite the Lagrangian density (\ref{A4}), 
\begin{eqnarray}
\label{A6}
\mathcal{L} =&   - \lambda_{(\Lambda)} 
+ \lambda_{(\alpha)} R + \lambda_{(\beta)} R^2 
+ \lambda_{(\gamma)} R_{\mu\nu} R^{\mu\nu} 
+ \lambda_{(\delta)} R_{\mu\nu\rho\sigma} R^{\mu\nu\rho\sigma} \nonumber \\
& + \partial_\mu \lambda_{(\Lambda)} 
\partial^\mu \varphi_{(\Lambda)}  
 - \partial_\mu b_{(\Lambda)} \partial^\mu c_{(\Lambda)} 
+ \partial_\mu \lambda_{(\alpha)} \partial^\mu \varphi_{(\alpha)}  
 - \partial_\mu b_{(\alpha)} \partial^\mu c_{(\alpha)} \nonumber \\
& + \partial_\mu \lambda_{(\beta)} \partial^\mu \varphi_{(\beta)}  
 - \partial_\mu b_{(\beta)} \partial^\mu c_{(\beta)} 
+ \partial_\mu \lambda_{(\gamma)} \partial^\mu \varphi_{(\gamma)}  
 - \partial_\mu b_{(\gamma)} \partial^\mu c_{(\gamma)} 
+ \mu \partial_\mu \lambda_{(\delta)} \partial^\mu \varphi_{(\delta)}  
 - \partial_\mu b_{(\delta)} \partial^\mu c_{(\delta)} \, , 
\end{eqnarray}
which tells that we can absorb the divergences into the redefinition of 
$\lambda_{(i)}$, 
$\left(i=\Lambda,\alpha,\beta,\gamma,\delta\right)$ and the divergences 
becomes irrelevant for the dynamics.\footnote{
In the model which includes the terms given by the square of curvatures 
in the Lagrangian density, 
there generally appear the massive scalar mode and/or the massive spin 2 mode. 
The latter is a ghost  and violates the unitarity. 
These modes do not appear only in the case that the curvature square terms are given 
by the Gauss-Bonnet combination. 
The massive scalar mode and the massive spin 2 mode can be regarded as composite modes, 
which do not appear in the perturbation by the definition of quantum field theory. 
Therefore we may need to renormalize the curvature square terms to be the Gauss-Bonnet 
combination. 
In the Lagrangian density (\ref{A6}), the evolution of the scalar fields $\lambda_{(i)}$ 
($i=\beta,\gamma,\delta$) is given by the common equations, 
$\nabla^\mu \partial_\mu \lambda_{(i)}=0$ as clear from the Lagrangian density (\ref{A6}). 
Therefore if we choose the initial condition or boundary condition so that $\lambda_{(i)}$'s become
the Gauss-Bonnet combination, the combination is preserved in whole space-time and the ghost does 
not appear. 
} 
The Lagrangian density (\ref{A6}) is  also invariant under 
the following BRS transformations
\begin{equation}
\label{A7}
\delta \lambda_{(i)} = \delta c_{(i)} = 0\, , \quad 
\delta \varphi_{(i)}i = \epsilon c \, , \quad 
\delta b_{(i)} = \epsilon \left( \lambda_{(i)} - \lambda_{(i)0} \right)
\, , 
\quad \left(i=\Lambda,\alpha,\beta,\gamma,\delta\right)\, ,
\end{equation}
where $\lambda_{(i)0}$'s satisfy the equation, 
\begin{equation}
\label{lambda0general}
0 = \nabla^\mu \partial_\mu \lambda_{(i)0}\, ,
\end{equation}
as in (\ref{lambda0}). 
The Lagrangian density (\ref{A6}) is also given by the BRS transformation 
(\ref{A7}) with $\lambda_{(i)0}=0$, 
\begin{equation}
\label{A8}
\delta \left( \sum_{i=\Lambda,\alpha,\beta,\gamma,\delta} 
\left(- b_{(i)} \left( 1 + \nabla_\mu \partial^\mu \varphi_{(i)} 
\right) \right) \right)
= \epsilon \left( \mathcal{L} 
+ \left(\mbox{total derivative terms}\right) \right)\, .
\end{equation}
As mentioned, due to the quantum correction from the graviton, the 
divergences in infinite numbers of quantum corrections appear. 
Let $\mathcal{O}_i$ be possible gravitational operators then 
a further generalization of the Lagrangian density (\ref{A6}) is given by
\begin{equation}
\label{A11}
\mathcal{L} = \sum_i \left( \lambda_{(i)} \mathcal{O}_{(i)} 
+ \partial_\mu \lambda_{(i)} \partial^\mu \varphi_{(i)}  
 - \partial_\mu b_{(i)} \partial^\mu c_{(i)} \right) \, .
\end{equation}
Then all the divergences are absorbed into the redefinition of $\lambda_i$. 
The Lagrangian density (\ref{A11}) is invariant under the BRS transformation 
and given by the the BRS transformation of some quantity and therefore 
the model can be regarded as a topological field theory, again. 

\section{Cosmology in Extended Model \label{Sec4}}

By the arguments in the last sections, the problems of the divergences in the quantum 
theory might be solved but there is not any principle to determine the values 
of the observed cosmological constant and other coupling constants. 
The values could be determined by the initial conditions or the boundary conditions 
in the classical theory. 
Therefore it could be interesting to investigate the cosmology and 
specify the region of the initial conditions which could be consistent with 
the evolution of the observed universe. 
For the model (\ref{CCC7}), in \cite{Mori:2017dhe}, it has been shown that we 
need the fine-tuning for the initial conditions although the constraints on the 
conditions are relaxed a little bit. 

For simplicity, we consider the following reduced model, 
\begin{equation}
\label{A6A}
\mathcal{L} =  - \lambda_{(\Lambda)} 
+ \lambda_{(\alpha)} R + \partial_\mu \lambda_{(\Lambda)} 
\partial^\mu \varphi_{(\Lambda)}  
 - \partial_\mu b_{(\Lambda)} \partial^\mu c_{(\Lambda)} 
+ \partial_\mu \lambda_{(\alpha)} \partial^\mu \varphi_{(\alpha)}  
 - \partial_\mu b_{(\alpha)} \partial^\mu c_{(\alpha)} \, .
\end{equation}
In order to consider the cosmology, we assume 
$b_{(\Lambda)} = c_{(\Lambda)} = b_{(\alpha)} = c_{(\alpha)}=0$ 
because the ghost number should be conserved and superselection 
rule should hold. 
We assume that the space-time is given by the FRW universe with flat spacial part, 
\begin{equation}
\label{FRW}
ds^2 = - dt^2 + a(t)^2 \sum_{i=1}^3 \left( dx^i \right)^2 \, ,
\end{equation}
and we assume that all the scalar fields $\lambda_{(\Lambda)}$, $\varphi_{(\Lambda)} $, 
$\lambda_{(\alpha)}$, and $\varphi_{(\alpha)}$ only depend on the cosmological time $t$. 
Then the variation of the action with respect to the scalar fields and metric 
gives the following equations. 
\begin{eqnarray}
\label{A6A2}
& 0= 1 + \left( \frac{d^2 \varphi_{(\Lambda)}}{dt^2} 
+ 3 H \frac{d \varphi_{(\Lambda)}}{dt} \right) \, , \quad 
0= \frac{d^2 \lambda_{(\Lambda)}}{dt^2} 
+ 3 H \frac{d \lambda_{(\Lambda)}}{dt} \, , \\
\label{A6A3}
& 0= 12 H^2 + 6 \frac{d H}{dt} + \left( \frac{d^2 \varphi_{(\alpha)}}{dt^2} 
+ 3 H \frac{d \varphi_{(\alpha)}}{dt} \right) \, , \quad 
0= \frac{d^2 \lambda_{(\alpha)}}{dt^2} 
+ 3 H \frac{d \lambda_{(\alpha)}}{dt} \, , \\
\label{SCCP5A}
& 3 \lambda_{(\alpha)} H^2 
= -3 H \frac{d \lambda_{(\alpha)}}{dt}
+ \lambda_{(\Lambda)} - \frac{d\lambda_{(\Lambda)}}{dt} \frac{d\varphi_{(\Lambda)}}{dt} 
 - \frac{d\lambda_{(\alpha)}}{dt} \frac{d\varphi_{(\alpha)}}{dt} \, , \\
\label{SCCP6A}
& -   \lambda_{(\alpha)} \left( 3 H^2 + 2 \frac{dH}{dt} \right) 
= 2 \frac{d^2 \lambda_{(\alpha)}}{dt^2} + 3 H \frac{d \lambda_{(\alpha)}}{dt}
- \lambda_{(\Lambda)} - \frac{d\lambda_{(\Lambda)}}{dt} \frac{d\varphi_{(\Lambda)}}{dt} 
 - \frac{d\lambda_{(\alpha)}}{dt} \frac{d\varphi_{(\alpha)}}{dt}\, .
\end{eqnarray}
If $\lambda_{(\Lambda)}$ and $\lambda_{(\alpha)}$ are constant, 
\begin{equation}
\label{A6A7}
\lambda_{(\Lambda)} = \lambda_{(\Lambda)0} \, , \quad 
\lambda_{(\alpha)} = \lambda_{(\alpha)0} \, ,
\end{equation}
the second equations in (\ref{A6A2}) and (\ref{A6A3}) are satisfied. 
Then Eq.~(\ref{SCCP5A}) or (\ref{SCCP6A}) gives $H$ is also a constant,
\begin{equation}
\label{A6A8}
H=H_0 \equiv \sqrt{\frac{\lambda_{(\Lambda)0}}{3\lambda_{(\alpha)0}}}\, .
\end{equation}
Then a solution for the first equations in (\ref{A6A2}) and (\ref{A6A3}) is given by
\begin{equation}
\label{A6A9}
\varphi_{(\Lambda)} = - \frac{t}{3H_0} \, , \quad 
\varphi_{(\alpha)} =  - 4H_0 t \, .
\end{equation}
Because $H$ is a constant, we find that the de Sitter space-time is a solution of this model. 

We now consider the stability of the obtained solution describing the de Sitter space-time by considering 
the perturbation, 
\begin{align}
\label{pert}
& H=H_{0}+\delta H\, ,\quad 
\lambda_{(\Lambda)} = \lambda_{(\Lambda)0} + \delta \lambda_{(\Lambda)} \, , \quad 
\varphi_{(\Lambda)} =  - \frac{t}{3H_0} + \delta \varphi_{(\Lambda)} \, , 
\nonumber \\  
& \lambda_{(\alpha)} = \lambda_{(\alpha)0} + \delta \lambda_{(\alpha)} \, , \quad 
\varphi_{(\alpha)} =  - 4 H_0 t + \delta \varphi_{(\alpha)} \, .
\end{align}
Then we obtain the following perturbed equations, 
\begin{eqnarray}
\label{perteq1A}
0=& \delta\ddot\varphi_{(\Lambda)} + 3H_{0}\delta\dot\varphi_{(\Lambda)}
 - \frac{1}{3H_0} \delta H \, , \\
\label{perteq2A}
0=& \delta\ddot{\lambda}_{(\Lambda)}+3H_{0}\delta\dot{\lambda}_{(\Lambda)} \, , \\
\label{perteq1B}
0=& \delta\ddot\varphi_{(\alpha)} + 3H_{0}\delta\dot\varphi_{(\alpha)}
 + 20 H_0 \delta H + 6 \delta \dot H 
\, , \\
\label{perteq2B}
0=& \delta\ddot{\lambda}_{(\alpha)}+3H_{0}\delta\dot{\lambda}_{(\alpha)} \, , \\
\label{perteq3A}
6 \lambda_{(\alpha)0} H_{0}\delta H + 3 H_0^2 \delta \lambda_{(\alpha)}=& 
 - 3 H_0 \delta {\dot\lambda}_{(\alpha)} + \delta \lambda_{(\Lambda)}
+ \frac{1}{3H_0}\delta \dot{\lambda}_{(\Lambda)} \, .
\end{eqnarray}
By using (\ref{perteq2A}), (\ref{perteq2B}), and (\ref{perteq3A}), we obtain
\begin{equation}
\label{perteq3B}
\lambda_{(\alpha)0} \delta \dot H = H_0 \delta {\dot\lambda}_{(\alpha)} \, .
\end{equation}
By using (\ref{perteq3A}), we delete $\delta H$ in (\ref{perteq1A}) and (\ref{perteq1B}) 
and obtain 
\begin{eqnarray}
\label{perteq1A2}
0=& \delta\ddot\varphi_{(\Lambda)} + 3H_{0}\delta\dot\varphi_{(\Lambda)}
 - \frac{1}{18H_0^2 \lambda_{(\alpha)0}} 
\left( - 3 H_0^2 \delta \lambda_{(\alpha)}
 - 3 H_0 \delta {\dot\lambda}_{(\alpha)} + \delta \lambda_{(\Lambda)}
+ \frac{1}{3H_0}\delta \dot{\lambda}_{(\Lambda)}\right) \, , \\
\label{perteq1B2}
0=& \delta\ddot\varphi_{(\alpha)} + 3H_{0}\delta\dot\varphi_{(\alpha)}
+ \frac{10}{3 \lambda_{(\alpha)0}} 
\left( - 3 H_0^2 \delta \lambda_{(\alpha)}
 - \frac{6}{5} H_0 \delta {\dot\lambda}_{(\alpha)} + \delta \lambda_{(\Lambda)}
+ \frac{1}{3H_0}\delta \dot{\lambda}_{(\Lambda)} \right) \, .
\end{eqnarray}
We now define new variables $\delta\eta_{(\Lambda)}$ and 
$\delta\eta_{(\alpha)}$ by 
\begin{equation}
\label{etaA}
\delta\eta_{(\Lambda)} \equiv \delta\dot\lambda_{(\Lambda)} \, , 
\quad 
\delta\eta_{(\alpha)} \equiv \delta\dot\lambda_{(\alpha)} \, , 
\end{equation}
and we rewrite (\ref{perteq2A}), (\ref{perteq2A}), (\ref{perteq1A2}), and 
(\ref{perteq1B2})  as follows, 
\begin{eqnarray}
\label{perteq5A}
0=& \delta\dot{\eta}_{(\Lambda)}+3H_{0}\delta\eta_{(\Lambda)} \, , \\
\label{perteq5B}
0=& \delta\dot{\eta}_{(\alpha)}+3H_{0}\delta\eta_{(\alpha)} \, , \\
\label{perteq1A3}
0=& \delta\ddot\varphi_{(\Lambda)} + 3H_{0}\delta\dot\varphi_{(\Lambda)}
 - \frac{1}{18H_0^2 \lambda_{(\alpha)0}} 
\left( - 3 H_0^2 \delta \lambda_{(\alpha)}
 -3 H_0 \delta \eta_{(\alpha)} + \delta \lambda_{(\Lambda)}
+ \frac{1}{3H_0}\delta \eta_{(\Lambda)}\right) \, , \\
\label{perteq1B2}
0=& \delta\ddot\varphi_{(\alpha)} + 3H_{0}\delta\dot\varphi_{(\alpha)}
+ \frac{10}{3 \lambda_{(\alpha)0}} 
\left( - 3 H_0^2 \delta \lambda_{(\alpha)}
 - \frac{6}{5} H_0 \delta \eta_{(\alpha)} + \delta \lambda_{(\Lambda)}
+ \frac{1}{3H_0}\delta \eta_{(\Lambda)} \right) \, .
\end{eqnarray}
Furthermore we define 
\begin{equation} 
\label{A6A10}
\delta \varphi \equiv \delta\dot\varphi_{(\Lambda)} 
+ \frac{1}{60 h_0^2} \delta\dot\varphi_{(\alpha)} \, ,
\end{equation}
Eqs.~(\ref{perteq1A3}) and (\ref{perteq1B2}) give 
\begin{equation}
\label{A6A10}
0 = \delta\dot\varphi + 3H_{0}\delta\varphi
 - \frac{\delta \eta_{(\alpha)}}{10H_0 \lambda_{(\alpha)0}} \, .
\end{equation} 
We now write Eqs.~(\ref{etaA}), (\ref{perteq5A}), (\ref{perteq5B}), 
(\ref{perteq1A3}), and (\ref{A6A10}) in a matrix form, 
\begin{equation}
\label{A6A11}
0 = \left( \frac{d}{dt} + A \right) 
\left( \begin{array}{c} 
\delta\eta_{(\Lambda)} \\ \delta\eta_{(\alpha)} \\
\delta \lambda_{(\Lambda)} \\ \delta \lambda_{(\alpha)} \\
\delta \varphi  \\ \delta\dot\varphi_{(\Lambda)} 
\end{array} \right) \, , \qquad 
A \equiv \left( \begin{array}{cccccc}
3H_0 & 0 & 0 & 0 & 0 & 0 \\
0 & 3H_0 & 0 & 0 & 0 & 0 \\
1 & 0 & 0 & 0 & 0 & 0 \\
0 & 1 & 0 & 0 & 0 & 0 \\
0 & \frac{1}{10H_0 \lambda_{(\alpha)0}} & 0 & 0 & 3H_0 & 0 \\
 - \frac{1}{54 H_0^3 \lambda_{(\alpha)0}} & {1}{6 H_0 \lambda_{(\alpha)0}} &
 - \frac{1}{18 H_0^2 \lambda_{(\alpha)0}} & \frac{1}{6 \lambda_{(\alpha)0}} & 
0 & 3 H_0 \end{array} \right) \, .
\end{equation}
If there is negative eigenvalue in the matrix $A$, the solution describing the de Sitter space-time 
is unstable but as clear from the form of the matrix $A$, which is triangular, the eigenvalues are 
given by four $3H_0$'s and two $0$'s. 
Therefore the solution describing the de Sitter space-time is stable or at least quasi-stable as in the 
model (\ref{CCC7R}) with $\mathcal{L}_\mathrm{gravity}^{(0)} = \frac{R}{2\kappa^2}$, that 
is, the case of the Einstein gravity \cite{Mori:2017dhe}. 
The stability tells that the solution might describe the acceleratingly expanding universe at present.  

\section{Summary and Discussions \label{Sec5}} 

In summary, in the model (\ref{A6A}), the divergences in the cosmological constant and the gravitational 
constant coming from the quantum corrections may not affect the dynamics but there is no principle 
to determine the constants, which may correspond to $\lambda_{(\Lambda)0}$ and 
$\lambda_{(\alpha)0}$ in (\ref{A6A7}). 
These constants could be determined by the initial conditions or something else and therefore it 
could be interesting to investigate the cosmology by including the matter as in the model of 
\cite{Mori:2017dhe}. 
In the model (\ref{A11}), however, we need infinite numbers of the initial conditions, which might 
be unphysical but this problem might give any clue for the quantum gravity. 
In fact, the second equations in (\ref{A6A2}) and (\ref{A6A3}) describe the evolutions of the 
effective coupling constants $\lambda_{(\Lambda)}$ and $\lambda_{(\alpha)}$ with respect to 
the scale $a(t)$ as in the renormalization group equations. 

Finally we mention on the relation with the Weinberg no-go theorem \cite{Weinberg:1988cp}. 
In the paper, it was assumed that the system has translational invariance and $GL(4)$ invariance. 
Then it has been shown that we need the fine-tuning the parameters in order to obtain the vanishing 
or small cosmological constant. 
In the paper, the translational invariance was assumed even for the fields and therefore 
all the fields are constant. 
As clear from Eq.~(\ref{A6A9}), some of the scalar fields must depend on the time and not constants. 
Therefore the Weinberg no-go theorem cannot apply for the model in this paper although there 
might be a problem of the fine-tuning for the initial conditions. 

\acknowledgments{This work is supported (in part) by 
MEXT KAKENHI Grant-in-Aid for Scientific Research on Innovative Areas ``Cosmic
Acceleration''  (No. 15H05890).}

\end{document}